\documentclass[12pt]{article}
\newcommand{\sect}[1]{\setcounter{equation}{0}\section{#1}\indent}

\def\[{{[}}
\def\CD{{\cal D}}

\def\CP{{\cal P}}

\def\PP{{\mathbb P}}

\def\xis{\xi\hspace{-2mm}/}
\def\Xis{\Xi\hspace{-2.5mm}/}
\def\etas{\eta\hspace{-2.0mm}/}

\def\nn{\nonumber}
\def\1{\mathbb I}
\newcommand{\bea}{\begin{eqnarray}}
\newcommand{\eea}{\end{eqnarray}}

\newcommand{\p}[1]{(\ref{#1})}
\renewcommand{\nn}{\nonumber}

\newcommand{\pa}{\partial}

\usepackage{amsbsy,amssymb,latexsym}

\begin{document}
\thispagestyle{empty}
\begin{flushright}
KEK-TH-889\\
hep-th/0306009
\end{flushright}
\vspace{30mm}

\begin{center}
{\bf\Large
IIB PP-Waves
with Extra Supersymmetries
}

\vspace{20mm}
Makoto Sakaguchi
\vspace{10mm}

\textit{Theory Division,
          High Energy Accelerator Research Organization (KEK)\\
           1-1 Oho, Tsukuba, Ibaraki, 305-0801, JAPAN}\\
\vspace{5mm}
           \texttt{Makoto.Sakaguchi@kek.jp}
\end{center}

\vspace{40mm}

\begin{abstract}

We examine Killing spinor equations of
the general IIB pp-wave backgrounds,
which contain a scalar $H(x^m,x^-)$ in the metric
and a self-dual four-form $\xi(x^m,x^-)$
in the self-dual five-form flux.
Considering non-harmonic extra Killing spinors,
we find that 
if the backgrounds admit at least one extra Killing spinor
in addition to 16 standard Killing spinors,
backgrounds can be reduced to the form with
$H=A_{mn}(x^-)x^mx^n$ and $\xi(x^-)$,
modulo coordinate transformations.
We examine further the cases in which
the extra Killing spinors are characterized
by a set of Cartan matrices.
Solving Killing spinor equations,
we find IIB pp-wave backgrounds
which admit 18, 20, 24 and 32 Killing spinors.

\end{abstract}
\newpage

\sect{Introduction}
After the advent of the maximally supersymmetric
IIB pp-wave background
\cite{BFHP:A new maximally},
pp-wave backgrounds have attracted renewed interests
among other supergravity solutions.
It was shown \cite{Metsaev:Type IIB} that 
the Green-Schwarz superstring on the
maximally supersymmetric IIB pp-wave 
background
is exactly solvable in the light-cone gauge
and the full string spectrum has been obtained.
Considering the large $N$ limit
corresponding to the Penrose limit~\cite{Penrose},
AdS/CFT correspondence has been examined \cite{BMN}
beyond the supergravity level.

General pp-wave solutions of IIB supergravity
admit at least sixteen standard Killing spinors.
At a special point in the moduli space,
the background turns out to be
the maximally supersymmetric pp-wave
solution~\cite{BFHP:A new maximally}
which admits sixteen extra Killing spinors in addition to the sixteen
standard Killing spinors, and thus maximal thirty-two
supersymmetries.
This background is a Penrose limit of the AdS$_5\times S^5$
background~\cite{BFCP:Penrose limits},
and the corresponding super-isometry algebras 
are related \cite{HKS:IIB} by an In\"on\"u-Wigner (IW) contraction.
In addition to the cases with sixteen and thirty-two supersymmetries,
it has been shown that there exist pp-wave backgrounds
which admit
20, 24 and 28 Killing spinors
~\cite{CLP:Penrose,
CLP:M-theory pp-waves,
CHKW:Penrose limit of RG fixed points,
BJLM:penrose limits deformed pp-waves,
BR:Supergravity,
GPS:penrose limit and RG flow}.

For eleven-dimensional supergravity,
the maximally supersymmetric pp-wave solution
called the Kowalski-Glikman (KG) solution
was obtained in \cite{Kowalski-Glikman:vacuum, Chrusciel:The isometry}.
This background is a Penrose limit of the
AdS$_{4/7}\times S^{7/4}$ backgrounds
\cite{BFCP:Penrose limits},
and the corresponding super-isometry algebras
\cite{Figueroa-O'Farrill:Homogeneous}
are related \cite{HKS:Super-pp-wave} by IW contractions.
It has been shown that there exist non-maximally supersymmetric solutions
with 18, 20, 22, 24, 26 supersymmetries~\cite{
CLP:Penrose,
Michelson:Twisted,
CLP:M-theory pp-waves,
GH:pp-waves in 11-dimensions,
Michelson:A pp-wave}.
For the type-IIA supergravity theory, the maximally supersymmetric
pp-wave solution does not exist \cite{FP;Maximally}.
The non-maximally cases were found in~\cite{Michelson:Twisted,
CLP:M-theory pp-waves,
Michelson:A pp-wave,
BR:Supergravity,
SY:IIA,
SS:String}.
For the lower dimensions, the maximally supersymmetric pp-wave solutions
were found in~\cite{Meessen:A small note} for five- and six-dimensions,
and in~\cite{Kowalski-Glikman:Positive} for four-dimensions.

In \cite{OS},
a uniqueness of eleven-dimensional
pp-wave solutions with extra supersymmetries
was discussed.
Examining Killing spinor equations of the general eleven-dimensional
pp-wave backgrounds, which contain a scalar $H(x^m,x^-)$
in the metric
and a three-form $\xi(x^m,x^-)$ in the flux,
it was shown that
if the backgrounds admit at least one non-harmonic extra Killing spinor
characterized by mutually commuting projectors
in addition to the standard 16 Killing spinors,
backgrounds can be reduced to the form with $H=A_{mn}(x^-)x^mx^n$
and $\xi(x^-)$, modulo coordinate transformations.
One of main purpose
of this paper is
to prove the similar uniqueness theorem for IIB pp-waves.

We examine
Killing spinor equations of the IIB pp-wave background
with a self-dual Ramond-Ramond (R-R) five-form flux,
\begin{eqnarray}
ds^2&=&
 2dx^+dx^- +H(x^m,x^-)(dx^-)^2+(dx^m)^2,~~
\nonumber\\
F&=&
dx^-\wedge \xi(x^m,x^-),~~~~\xi=*_8\xi
\label{pp:IIB}
\end{eqnarray}
where $\xi$ is a self-dual four-form on the transverse
$\mathbb E^8$
spanned by $x^m$.
We show that the IIB pp-wave background~(\ref{pp:IIB})
is highly restricted if there is at least one non-harmonic
extra Killing spinor,
so that $H=A_{mn}(x^-)x^mx^n$ and $\xi(x^-)$,
modulo coordinate transformations.

We further examine Killing spinor equations
and consider two types of $\xi$.
One is related to the K\"ahler form of a Calabi-Yau four-fold with
$SU(4)$ holonomy,
and the other is the self-dual Cayley four-form
of $d=8$ Riemannian manifold with $Spin(7)$ holonomy.
The former case
can combine
Neveu-Schwarz-Neveu-Schwarz (NS-NS)  and R-R three-forms
and has been examined well.
Examining the latter case,
we find IIB pp-wave solutions
which have not been given
in the literature yet,
as long as we know.
Our solutions admit 18, 20, 24 and 32 supersymmetries.
The pp-wave solution with 32 supersymmetries
is shown to be related to the solution given in \cite{BFHP:A new maximally}
by a coordinate transformation.

This paper is organized as follows.
In the next section,
Killing spinor equations for the background~(\ref{pp:IIB})
is derived.
In section 3, we prove a uniqueness theorem which states that
$H(x^-,x^m)$ and $\xi(x^-,x^m)$ can be reduced to
$A_{mn}(x^-)x^mx^n$
and $\xi(x^-)$, respectively,
modulo coordinate transformations,
provided that the background admits at least one non-harmonic
extra Killing spinor
in addition to the standard sixteen spinors.
We further examine Killing spinor equations
for the cases in which extra Killing spinors
are charactereized by mutually commuting projectors in section 4.
IIB pp-wave solutions with extra supersymmetries are given in section 5.
The last section is devoted to a summary and discussions.

\sect{Killing spinor equations}
The general IIB pp-wave background we consider is
\begin{eqnarray}
ds^2&=&
 2dx^+dx^- +H(x^m,x^-)(dx^-)^2+(dx^m)^2,\\
F&=&dx^-\wedge \xi(x^m,x^-),~~~~\xi=*_8\xi
\label{background}
\end{eqnarray}
where both a scalar $H$ and a self-dual four-form $\xi$
on $\mathbb E^8$ spanned by $x^m$, are
functions of $x^-$ and $x^m$. 
This is a supergravity solution when
\begin{eqnarray}
\triangle H=
-\frac{32}{4!}\xi_{klmn}\xi^{klmn},
\label{sugra}
\end{eqnarray}
where $\triangle$ is the Laplacian on $\mathbb E^8$.
The frame one-forms defined by $ds^2=2e^+e^-+e^me^m$ are
\begin{eqnarray}
e^-=dx^-,~~~
e^+=dx^++\frac{1}{2}H(x^m,x^-)dx^-,~~~
e^m=dx^m
\end{eqnarray}
and thus the spin connection is
\begin{eqnarray}
w^+{}_m=\frac{1}{2}\partial_m H dx^-.
\end{eqnarray}
Killing spinor equations for general IIB backgrounds 
with a R-R five-form field strength $F_{M_1\cdots M_5}$ 
\begin{eqnarray}
&&\CD_M\varepsilon=(\nabla_M-\Omega_M)\varepsilon=0,
\nonumber\\&&\nabla_M=\partial_M
+\frac{1}{4}w_M^{ab}\Gamma_{ab},~~~
\Omega_M=-\frac{i}{24}F_{ML_1\cdots L_4}\Gamma^{L_1\cdots L_4},
\end{eqnarray}
reduce on this pp-wave background to
\begin{eqnarray}
\partial_+\varepsilon=0,~~~
\partial_-\varepsilon
 -\frac{1}{4}\partial_mH\Gamma^m\Gamma_+\varepsilon
 =\Omega_-\varepsilon,~~~
\partial_m\varepsilon=\Omega_m\varepsilon,
\end{eqnarray}
where
\begin{eqnarray}
\Omega_m=-i\Xis_m\Gamma_+,~~~
\Omega_-=-i\Xis,~~~
\Xis=\frac{1}{4!}\xi_{pqrs}\Gamma^{pqrs},~~~
\Xis_m=\frac{1}{3!}\xi_{mpqr}\Gamma^{pqr}.
\end{eqnarray}
It is convenient to introduce the eight-dimensional gamma matrices
$\gamma^m \in Spin(8)$ 
\begin{eqnarray}
\Gamma_0=\1_{16}\otimes i\sigma_2,~~~
\Gamma_9=\1_{16}\otimes\sigma_1,~~~
\Gamma_m=\gamma_m\otimes\sigma_3.
\label{8dim}
\end{eqnarray}
Defining the light-cone projection operator as
\begin{eqnarray}
\CP_\pm=\frac{1}{2}\Gamma_\pm\Gamma_\mp,~~~
\Gamma_\pm=\frac{1}{\sqrt{2}}(\Gamma_9\pm\Gamma_0),
\end{eqnarray}
the complex Weyl spinor $\varepsilon$
decomposes into
\begin{eqnarray}
\varepsilon=\left(
  \begin{array}{c}
    \varepsilon_+   \\
    \varepsilon_-   \\
  \end{array}
\right),~~~
\CP_+\varepsilon=\left(
  \begin{array}{c}
    \varepsilon_+   \\
    0   \\
  \end{array}
\right),~~~
\CP_-\varepsilon=\left(
  \begin{array}{c}
    0   \\
    \varepsilon_-   \\
  \end{array}
\right).
\end{eqnarray}
Defining the chirality projection operator as
\begin{eqnarray}
h_+=\frac{1}{2}(1+\Gamma_{11}),~~~
\Gamma_{11}=\Gamma_{012\cdots 9}=\gamma_{12\cdots 8}\otimes\sigma_3,
\end{eqnarray}
the positive chirality condition, $h_+\varepsilon=\varepsilon$,
implies
that
\begin{eqnarray}
\gamma_{12\cdots 8}\varepsilon_\pm=\pm\varepsilon_\pm,
\label{SD}
\end{eqnarray}
which halves
the 32 complex components of $\varepsilon$
into 16 complex components.
$\varepsilon_+$ is called the \textit{standard} Killing spinor
which exists on the general pp-wave backgrounds,
while $\varepsilon_-$ is the \textit{extra} Killing spinor.
Defining $\xis$ and $\xis_m$ as
\begin{eqnarray}
\xis=\frac{1}{4!}\xi_{lmnp}\gamma^{lmnp},~~~
\xis_m=\frac{1}{3!}\xi_{mnpq}\gamma^{npq},~~~
\Xis\equiv\xis\otimes \1,~~~
\Xis_m\equiv \xis_m\otimes \sigma_3,
\end{eqnarray}
the Killing spinor equations
are expressed as
\begin{eqnarray}
&&\partial_+\varepsilon_+=0,\label{1}\\
&&\partial_-\varepsilon_+
 -\frac{\sqrt{2}}{4}\partial_mH\gamma_{m}\varepsilon_-
 =-i\xis\varepsilon_+,\label{2}\\
&&\partial_m\varepsilon_+=
 -i\sqrt{2}\xis_m\varepsilon_-,\label{3}\\
&&\partial_+\varepsilon_-=0,\label{4}\\
&&\partial_-\varepsilon_-=0,\label{5}\\
&&\partial_m\varepsilon_-=0,\label{6}
\end{eqnarray}
where we have used the fact $\xis\varepsilon_-=0$,
which follows from the self-duality property of $\xis$, 
(\ref{background}),
and the positive chirality property of $\varepsilon_-$, 
(\ref{SD}).  

\sect{Uniqueness}
In this section, we examine Killing spinor equations
(\ref{1})--(\ref{6})
and derive conditions on $H$ and $\xi$,
providing that there exists at least one extra Killing spinor.

It follows from eqns. (\ref{4}), (\ref{5}) and (\ref{6}) that
$\varepsilon_-$ is a constant spinor.
Eqn. (\ref{1}) implies that
$\varepsilon_+$ must be independent of $x^+$.
Acting $\gamma^m$ on (\ref{3}),
one finds
\begin{eqnarray}
\gamma^m\partial_m\varepsilon_+=0
\label{d}
\end{eqnarray}
because $\gamma^m\xis_m\varepsilon_-=4\xis\varepsilon_-=0$.
Further acting $\gamma^n\partial_n$ on this equation, we find that
$\varepsilon_+$ obeys the Laplace equation
\begin{eqnarray}
\partial_m\partial^m\varepsilon_+=0.
\label{dd}
\end{eqnarray}
It follows that $\varepsilon_+$ is linear in $x^m$ at most,
up to a harmonic function.
In this paper, we concentrate on the non-harmonic function
part.\footnote{
PP-wave backgrounds which admit harmonic Killing spinors
were extensively discussed in \cite{MM:Strings}.
}
We can thus write $\varepsilon_+$ as
\begin{eqnarray}
\varepsilon_+=\varepsilon_0(x^-)+\varepsilon_m(x^-)x^m
\end{eqnarray}
where $\varepsilon_0$ and $\varepsilon_m$ are functions of
$x^-$ only
and eqn.(\ref{3}) becomes
\begin{eqnarray}
\varepsilon_m=-i\sqrt{2}\xis_m\varepsilon_-.
\label{3'}
\end{eqnarray}
Because one can show that
\begin{eqnarray}
\xis_m\varepsilon_-=-\frac{1}{2}\xis\gamma_m\varepsilon_-
\end{eqnarray}
this equation becomes
\begin{eqnarray}
\varepsilon_m=i\frac{\sqrt{2}}{2}\xis\gamma_m\varepsilon_-.
\end{eqnarray}
Since $\varepsilon_m$ depends only on $x^-$
and $\varepsilon_-$ is a constant spinor,
$\xis\gamma_m\PP$ must depend only on $x^-$ for all $m$,
where $\PP$ is a projection operator to
non-trivial extra Killing spinors
defined by
\begin{eqnarray}
\PP\varepsilon_-=\varepsilon_-,~~~
\overline{\PP}\varepsilon_-=0,~~~
\PP+\overline{\PP}=\1.
\end{eqnarray}
Because $\PP$ is constructed in terms of gamma matrices
as will be seen in the next section,
the term $\xis\gamma_m\PP$ 
is equal either to $\xis\PP\gamma_m$ for a certain set of $m$
or $\xis\overline{\PP}\gamma_m$ for the rest of $m$. 
The former (latter) case implies that
$\xis\PP$ ($\xis\overline{\PP}$)
depends only on $x^-$,
and thus $\xis$ depends only on $x^-$.
It follows from (\ref{2}) that $\partial_mH$
must be linear in $x^m$ at most,
and thus $H$ can be written as
$H=f(x^-)+g_m(x^-)x^m+A_{mn}(x^-)x^mx^n$,
where $f$, $g_m$ and $A_{mn}$ are functions of $x^-$.
As was done in \cite{OS}, one can show  that
$f(x^-)$ and $g_m(x^-)$ can be absorbed by a redefinition
\begin{eqnarray}
x^+ = y^+ - F(x^-) - G_m(x^-)y^m,~~~
x^m = y^m - H^m(x^-),
\label{redefinition}
\end{eqnarray}
where $F$, $G$ and $H$ satisfy
\begin{eqnarray}
&&-\pa F+\frac{1}{2}f-g_mH^m+\frac{1}{2}A_{mn}H^mH^n
 + \frac{1}{2}(\pa H^m)^2=0,\\
&& -\pa G_m +\frac{1}{2}g_m -A_{mn}H^n =0,\\
&&-G_m -\pa H_m=0,
\end{eqnarray}
and the line element reduces to
\begin{eqnarray}
ds^2&=& 2dy^+dx^- + A_{mn}(x^-) y^m y^n (dx^-)^2 +(dy^m)^2.
\end{eqnarray}
The transformation (\ref{redefinition}) does not affect
$F=dx^-\wedge \xi(x^-)$.
In summary, we have shown that IIB pp-wave backgrounds
which admit extra Killing spinors
must be of the form
\begin{eqnarray}
ds^2=
 2dx^+dx^- +A_{mn}(x^-)x^mx^n(dx^-)^2+(dx^m)^2,~~~
F=dx^-\wedge \xi(x^-),
\label{pp1}
\end{eqnarray}
modulo coordinate transformations.\footnote{
It should be noted again that
we are considering non-harmonic extra Killing spinors.
}

$A_{mn}(x^-)$ and $\xi(x^-)$
are restricted by (\ref{2}).
On the background (\ref{pp1}),
eqn.(\ref{2}) reduces to
\begin{eqnarray}
\partial_-\varepsilon_m
-\frac{\sqrt{2}}{2}A_{mn}\gamma^n\varepsilon_-
 =-i\xis\varepsilon_m,
\end{eqnarray}
which becomes, substituting (\ref{3'}) into this equation,
\begin{eqnarray}
[i\partial_-\xis_{(m)}-\xis^2_{(m)}
+\etas_{(m)}
-A_{m}]\varepsilon_-
\equiv D_{(m)}\varepsilon_-=0,
\label{KSE}
\end{eqnarray}
where
\begin{eqnarray}
\gamma^m\xis_{(m)}\equiv \xis\gamma^m,~~~
\etas_{(m)}\equiv -\sum_{n=1}^8A_{mn}\gamma^{mn},~~~
A_m\equiv A_{mm}. 
\end{eqnarray}
We examine this condition in the next section
for the case in which $D_{(m)}$
is expanded solely in terms of mutually commuting projectors.

\sect{$D_{(m)}$ expanded in mutually commuting projectors}
For non-maximally supersymmetric backgrounds,
$D_{(m)}$ must be a linear combination of projection operators.
We restrict the study to the case in which
$D_{(m)}$
is expanded \textit{solely} in terms of mutually commuting projectors.\footnote{
More general cases are studied in \cite{BMO}.
We thank the authors for explanation of their work.}
These projection operators are composed of
a set of Cartan matrices,
$H_I$, as
$
P_I=\frac{1}{2}(\1+H_I)
$.
Among infinitely many Cartan matrices,
we consider Cartan matrices which are monomials
of gamma-matrices, such as $H_I=\gamma^{[N_I]}$
where $\gamma^{[N]}$ is an $N$-th antisymmetrized product
of gamma matrices.
One finds that mutually commuting matrices $\gamma^{[N]}$ must
share a definite number of indices. We indicate the number of common
indices shared among two of matrices below.
\begin{eqnarray}
  \begin{array}{c|cc}
       & \gamma^{n_1...n_{2j}}   
       & \gamma^{n_1...n_{2j+1}}   \\
\hline
\gamma^{m_1...m_{2i}} &0,2,...,min(2i,2j)    
                      &0,2,...,min(2i,2j+1)    \\
\gamma^{m_1...m_{2i+1}} &0,2,...,min(2i+1,2j)    
                        &1,3,...,min(2i+1,2j+1)    \\
  \end{array}
\end{eqnarray}
The projection operators must commute with
the chirality projection operator,
so that matrices must commute with $\gamma^{12...8}$.
We find that there are two sets of mutually commuting matrices;
one is
\begin{eqnarray}
&&
\gamma^{12},
\gamma^{34},
\gamma^{56},
\gamma^{78}
\nonumber\\&&
\gamma^{1234},
\gamma^{3456},
\gamma^{5678},
\gamma^{1256},
\gamma^{1278},
\gamma^{3478}
\nonumber\\&&
\gamma^{123456},
\gamma^{123478},
\gamma^{125678},
\gamma^{345678},
\nonumber\\&&
\gamma^{12345678},
\label{matrices-1}
\end{eqnarray}
and the other is
\begin{eqnarray}
&&
\gamma^{1238},
\gamma^{1458},
\gamma^{1678},
\gamma^{2468},
\gamma^{2578},
\gamma^{3478},
\gamma^{3568},
\nonumber\\&&
\gamma^{4567},
\gamma^{2367},
\gamma^{2345},
\gamma^{1357},
\gamma^{1346},
\gamma^{1256},
\gamma^{1247},
\nonumber\\&&
\gamma^{12345678}.
\label{matrices-2}
\end{eqnarray}
The first set (\ref{matrices-1})
can be related to the K\"ahler form $J$ of
a Calabi-Yau four-fold
with $SU(4)$ holonomy.
The K\"ahler form $J$ is covariantly constant $dJ=0$.
The terms in the first line of (\ref{matrices-1})
are the constituents
of $J$, and those in the second, third and fourth lines
are the constituents of
 $J\wedge J$, $J\wedge J\wedge J$ and
 $J\wedge J\wedge J\wedge J$,
respectively. 
On the other hand, the second set (\ref{matrices-2})
can be related to the self-dual Cayley four-form $\Psi$
of $d=8$ Riemannian manifold with $Spin(7)$ holonomy,
which is
covariantly constant $d\Psi=0$.
The terms in the first and second lines of
(\ref{matrices-2})
are the constituents
of $\Psi$,
and those in the third line is the constituent of
$\Psi\wedge \Psi$.
The relation between covariantly constant forms
and special holonomy groups
is well known \cite{Wang}.
We find that there are three other possibilities to construct
the self-dual four-form from invariant forms,
which are shown to reduce to the self-dual Cayley four-form $\Psi$.
One is to use
the associative three-form $\phi$
of $d=7$ Riemannian manifold with $G_2$
holonomy and to make a wedge product of $\phi$
and a one-form, say $e^8$.
The self-dual four-form is obtained 
by adding eight-dimensional Hodge dual, $*_8$, of $\phi\wedge e^8$
as
$\phi\wedge e^8+*_8(\phi\wedge e^8)=\phi\wedge e^8+*_7\phi$.
This is nothing but the self-dual Cayley four-form $\Psi$.
Another is to use the seven-dimensional Hodge dual, $*_7$, of
$\phi$.
Again, the self-dual four-form turns out to be the
self-dual Cayley four-form $\Psi$,
because $*_7\phi+*_8(*_7\phi)=*_7\phi+\phi\wedge e^8$.
The third is to use
the holomorphic $(4,0)$-form $\Omega$
of a Calabi-Yau four-fold with $SU(4)$ holonomy.
A self-dual four-form can be constructed as the real or
imaginary part of
$\Omega$, which turns out to be a part of the self-dual
Cayley four-form $\Psi$.

Now $D_{(m)}$ is a linear combination of projection operators,
and thus must be constructed from these matrices.
Noting that $\xi=*_8\xi$,
$\xis$ turns out to be of the form
\begin{eqnarray}
\xis=
 a_1(\gamma^{1234}+\gamma^{5678})
 +a_2(\gamma^{3456}+\gamma^{1278})
 +a_3(\gamma^{1256}+\gamma^{3478})
\label{xis1}
\end{eqnarray}
for the former set~\p{matrices-1},
while
\begin{eqnarray}
\xis&=&
  b_1(\gamma^{4567}+\gamma^{1238})
 +b_2(\gamma^{2367}+\gamma^{1458})
 +b_3(\gamma^{2345}+\gamma^{1678})
 +b_4(\gamma^{1357}+\gamma^{2468})
 \nonumber\\&&
 +b_5(\gamma^{1346}+\gamma^{2578})
 +b_6(\gamma^{1256}+\gamma^{3478})
 +b_7(\gamma^{1247}+\gamma^{3568})
\label{xis2}
\end{eqnarray}
for the latter set~\p{matrices-2}.
$A_{mn}$ is restricted to be non-trivial only when
$m=n$ and $(m,n)=(1,2),~(3,4),~(5,6),~(7,8)$ for
(\ref{matrices-1}),
while only when $m=n$ for (\ref{matrices-2}).
In the former case,
$\xis_{(m)}$ for~\p{xis1} and $\etas_{(m)}$ are expressed as
\begin{eqnarray}
\xis_{(m)}&=&
   \alpha_1^m(\gamma^{1234}
-\gamma^{5678})
 + \alpha_2^m(\gamma^{3456}
-\gamma^{1278})
 + \alpha_3^m(\gamma^{1256}
-\gamma^{3478}),\nonumber\\
\etas_{(m)}&=&
\mu_1^m\gamma^{12}+\mu_2^m\gamma^{34}
 +\mu_3^m\gamma^{56}+\mu_4^m\gamma^{78},
\label{xim1}
\end{eqnarray}
where
\begin{eqnarray}
\alpha_1&=&(-a_1,-a_1,-a_1,-a_1,+a_1,+a_1,+a_1,+a_1)
,\nn\\
\alpha_2&=&(+a_2,+a_2,-a_2,-a_2,-a_2,-a_2,+a_2,+a_2)
,\nn\\
\alpha_3&=&(-a_3,-a_3,+a_3,+a_3,-a_3,-a_3,+a_3,+a_3),
\nonumber\\
\mu_i^{2i-1}&=&-\mu_i^{2i} =A_{2i-1~2i},~~~i=1,2,3,4.
\label{alpha}
\end{eqnarray}
In the latter case,
$\xis_{(m)}$ for~\p{xis2} and $\etas_{(m)}$ are expressed as
\begin{eqnarray}
\xis_{(m)}&=&
   \beta_1^m  (\gamma^{4567}-\gamma^{1238})
 + \beta_2^m  (\gamma^{2367}-\gamma^{1458})
 + \beta_3^m  (\gamma^{2345}-\gamma^{1678})
 + \beta_{4}^m(\gamma^{1357}-\gamma^{2468})
 \nonumber\\&&
 + \beta_{5}^m(\gamma^{1346}-\gamma^{2578})
 + \beta_{6}^m(\gamma^{1256}-\gamma^{3478})
 + \beta_{7}^m(\gamma^{1247}-\gamma^{3568})
 \nonumber\\
\etas_{(m)}&=&0
\label{xim2}
\end{eqnarray}
where
\begin{eqnarray}
\beta_{1}&=& (+b_1,+b_1,+b_1,-b_1,-b_1,-b_1,-b_1,+b_1),\nonumber\\
\beta_{2}&=& (+b_2,-b_2,-b_2,+b_2,+b_2,-b_2,-b_2,+b_2),\nn\\
\beta_{3}&=& (+b_3,-b_3,-b_3,-b_3,-b_3,+b_3,+b_3,+b_3),\nn\\
\beta_{4}&=& (-b_4,+b_4,-b_4,+b_4,-b_4,+b_4,-b_4,+b_4),\nn\\
\beta_{5}&=& (-b_5,+b_5,-b_5,-b_5,+b_5,-b_5,+b_5,+b_5),\nn\\
\beta_{6}&=& (-b_6,-b_6,+b_6,+b_6,-b_6,-b_6,+b_6,+b_6),\nn\\
\beta_{7}&=& (-b_7,-b_7,+b_7,-b_7,+b_7,+b_7,-b_7,+b_7).
\label{beta}
\end{eqnarray}
In the following two subsections,
we examine (\ref{KSE}) for these two cases in turn.
\subsection{the case~\p{xim1}}

In this case, all the matrices (\ref{matrices-1}) can be
constructed as products of four matrices, $\gamma^{12}$, $\gamma^{34}$,
$\gamma^{56}$ and $\gamma^{78}$. From these matrices, we make
four rank-8 projection operators
\begin{eqnarray}
&&P_1=\frac{1}{2}(\1+i\gamma^{12}),~~~
P_2=\frac{1}{2}(\1+i\gamma^{34}),
\nonumber\\&&
P_3=\frac{1}{2}(\1+i\gamma^{56}),~~~
P_4=\frac{1}{2}(\1+i\gamma^{78}),
\label{P1}
\end{eqnarray}
which satisfy
\begin{eqnarray}
P_A^2=P_A,~~~
P_AP_B=P_BP_A,~~~
A,B=1,2,3,4.
\end{eqnarray}
We rewrite (\ref{KSE}) in terms of these projection operators.
Noting $\gamma^{12..8}\varepsilon_-=-\varepsilon_-$,
one finds that
\begin{eqnarray}
\xis_{(m)}\varepsilon_-&=&
(   2\alpha_1^m\gamma^{1234}
 + 2\alpha_2^m\gamma^{3456}
 + 2\alpha_3^m\gamma^{1256}
)\varepsilon_-,
\nonumber\\
\etas_{(m)}\varepsilon_-&=&
(
\mu_1^m\gamma^{12}
+\mu_2^m\gamma^{34}
+\mu_3^m\gamma^{56}
+\mu_4^m\gamma^{123456}
)\varepsilon_-.
\label{xim1 on e}
\end{eqnarray}
Because
\begin{eqnarray}
&&
\gamma^{1234}=-(2 P_1-\1)(2P_2-\1),~~~
\gamma^{3456}=-(2P_2-\1)(2P_3-\1),~~~
\nonumber\\&&
\gamma^{1256}=-(2P_1-\1)(2P_3-\1),~~~
\gamma^{12}=-i(2P_1-\1),~~~
\gamma^{34}=-i(2P_2-\1),
\nonumber\\&&
\gamma^{56}=-i(2P_3-\1),~~~
\gamma^{123456}=i(2P_1-\1)(2P_2-\1)(2P_3-\1),
\end{eqnarray}
$\xis_{(m)}\varepsilon_-$ and $\etas_{(m)}\varepsilon_-$
in (\ref{xim1 on e})
can be rewritten as
\begin{eqnarray}
\xis_{(m)}\varepsilon_-&=&\Big[~
-2(\alpha^m_1+\alpha^m_2+\alpha^m_3)
+4(\alpha^m_1+\alpha^m_3)P_1
+4(\alpha^m_1+\alpha^m_2)P_2
+4(\alpha^m_2+\alpha^m_3)P_3
\nonumber\\&&~~
-8\alpha^m_1P_1P_2
-8\alpha^m_3P_1P_3
-8\alpha^m_2P_2P_3
~~\Big]\varepsilon_-,
\nonumber\\
\etas_{(m)}\varepsilon_-&=&
\Big[~
i(\mu_1^m+\mu_2^m+\mu_3^m-\mu_4^m)\1
-2i(\mu^m_1-\mu^m_4)P_1
-2i(\mu^m_2-\mu^m_4)P_2
\\&&~~
-2i(\mu^m_3-\mu^m_4)P_3
-4i\mu^m_4P_1P_2
-4i\mu^m_4P_1P_3
-4i\mu^m_4P_2P_3
+8i\mu^m_4P_1P_2P_3
~\Big]\varepsilon_-.\nonumber
\end{eqnarray}
Substituting these into (\ref{KSE}) yields
\begin{eqnarray}
&\Bigl[&
\Big(
-2i\partial_-(
\alpha^m_1+\alpha^m_2+\alpha^m_3
)^2
+i(\mu_1^m+\mu_2^m+\mu_3^m-\mu_4^m)
-4(\alpha^m_1+\alpha^m_2+\alpha^m_3
)^2
-A_m
\Big)~\1
\nonumber\\&&
+\Big(
 4i\partial_-(
 \alpha^m_1+\alpha^m_3
 )
-2i(\mu^m_1-\mu^m_4)
 +16\alpha^m_2(\alpha^m_1+\alpha^m_3)
\Big)~P_1
\nonumber\\&&
+\Big(
 4i\partial_-(
 \alpha^m_1+\alpha^m_2
 )
-2i(\mu^m_2-\mu^m_4)
 +16\alpha^m_3(\alpha^m_1+\alpha^m_2)
\Big)~P_2
\nonumber\\&&
+\Big(
 4i\partial_-(
 \alpha^m_2+\alpha^m_3
 )
-2i(\mu^m_3-\mu^m_4)
 +16\alpha^m_1(\alpha^m_2+\alpha^m_3)
\Big)~P_3
\nonumber\\&&
+\Big(
 -8i\partial_-\alpha^m_1
-4i\mu^m_4
 -32\alpha^m_2\alpha^m_3
\Big)~P_1P_2
+\Big(
 -8i\partial_-\alpha^m_2 
-4i\mu^m_4
 -32\alpha^m_1\alpha^m_3
\Big)~P_2P_3
\nonumber\\&&
+\Big(
-8i\partial_-\alpha^m_3
-4i\mu^m_4
 -32\alpha^m_1\alpha^m_2
\Big)~P_1P_3
+\Big(
8i\mu^m_4
\Big)~P_1P_2P_3
~~~~
\Bigr]\varepsilon_-=0.
\label{KSE1}
\end{eqnarray}
In order to see which Killing spinor survives,
it is convenient to introduce rank-1 projection
operators of a 16-component
spinor onto the $I$-th component:
\begin{eqnarray}
\PP_I=diag(0,\ldots,0,\stackrel{I}{1},0,\ldots,0).
\end{eqnarray}
The rank-8 projection operators $P_A$ can
then be expressed in terms of
these rank-1 projection operators as
\begin{eqnarray}
&&P_1=\sum_{I=1,2,..,8}\PP_I,~~~
P_2=\sum_{I=1,2,3,4,9,10,11,12}\PP_I,\nonumber\\&&
P_3=\sum_{I=1,2,5,6,9,10,13,14}\PP_I,~~~
P_4=\sum_{I=1,3,5,7,9,11,13,15}\PP_I.
\label{PP1}
\end{eqnarray}
The chirality condition,
$\gamma^{12...8}\varepsilon_-=-\varepsilon_-$,
reduces to 
\begin{eqnarray}
\frac{1}{2}(\gamma^{12...8}+\1)\varepsilon_-&=&
\frac{1}{2}\Big((2P_1-\1)(2P_2-\1)(2P_3-\1)(2P_4-\1)+\1\Big)
\nonumber\\
&=&
\sum_{I=1,4,6,7,10,11,13,16}\PP_I\varepsilon_-=0
\end{eqnarray}
and thus non-trivial spinors are 
$\PP_I\varepsilon_-$ with $I=
2,3,5,8,9,12,14$ and $15$.
Taking this into account,
(\ref{KSE1}) becomes
\begin{eqnarray}
&
\Bigl[&
\Big(
-2i\partial_-(
\alpha^m_1+\alpha^m_2+\alpha^m_3
)
+i(\mu_1^m+\mu_2^m+\mu_3^m-\mu_4^m)
-4(\alpha^m_1+\alpha^m_2+\alpha^m_3
)^2
-A_m
\Big)~\1
\nonumber\\&&
+\Big(
-2i(\mu^m_1
+\mu^m_2
+\mu^m_3
-\mu^m_4)
\Big)~\PP_2
\nonumber\\&&
+\Big(
4i\partial_-(\alpha^m_2+\alpha^m_3)
-2i(\mu_1^m+\mu_2^m)
+16\alpha^m_1(\alpha^m_2+\alpha^m_3)
\Big)~\PP_3
\nonumber\\&&
+\Big(
4i\partial_-(\alpha^m_1+\alpha^m_2)
-2i(\mu_1^m+\mu_3^m)
+16\alpha^m_3(\alpha^m_1+\alpha^m_2)
\Big)~
\PP_5
\nonumber\\&&
+\Big(
4i\partial_-(\alpha^m_1+\alpha^m_3)
-2i(\mu_1^m-\mu_4^m)
+16\alpha^m_2(\alpha^m_1+\alpha^m_3)
\Big)~\PP_8
\nonumber\\&&
+\Big(
4i\partial_-(\alpha^m_1+\alpha^m_3)
-2i(\mu_2^m+\mu_3^m)
+16\alpha^m_2(\alpha^m_1+\alpha^m_3)
\Big)~
\PP_9
\nonumber\\&&
+\Big(
4i\partial_-(\alpha^m_1+\alpha^m_2)
-2i(\mu_2^m-\mu_4^m)
+16\alpha^m_3(\alpha^m_1+\alpha^m_2)
\Big)~\PP_{12}
\nonumber\\&&
+\Big(
4i\partial_-(\alpha^m_2+\alpha^m_3)
-2i(\mu_3^m-\mu_4^m)
+16\alpha^m_1(\alpha^m_2+\alpha^m_3)
\Big)~
\PP_{14}
~~
\Bigr]\varepsilon_-=0.
\end{eqnarray}
It follows that the coefficient of $\1$ in this equation
must vanish in order to get a complex extra Killing spinor,
$\PP_{15}\varepsilon_-$.
Because $\alpha_i^m$, $\mu_i^m$ and $A_m$ are real, 
this implies two equations
\begin{eqnarray}
&&-2\partial_-(\alpha^m_1+\alpha^m_2+\alpha^m_3)
 +(\mu_1^m+\mu_2^m+\mu_3^m-\mu_4^m)=0,
\label{partial1}\\
&&A_m=-4(\alpha^m_1+\alpha^m_2+\alpha^m_3)^2.
\label{A_m1}
\end{eqnarray}
Without loss of generality,
we can take these equations as the conditions for the
existence of a complex extra Killing spinor,
because the condition for another spinor to be
a complex extra Killing spinor
is simply obtained by changing signs of
$\alpha_i^m$ and $\mu_i^m$.
The first equation (\ref{partial1}) leads to
\begin{eqnarray}
  \begin{array}{l}
-2\partial_-(\alpha^{2n-1}_1+\alpha^{2n-1}_2+\alpha^{2n-1}_3)
 +\mu_n^{2n-1}=0,\\
-2\partial_-(\alpha^{2n}_1+\alpha^{2n}_2+\alpha^{2n}_3)
 +\mu_n^{2n}=0,\\
  \end{array}
~~~~n=1,2,3,4,
\end{eqnarray}
from which we find that $\mu_i^m=0$,
so that $A_{12}=A_{34}=A_{56}=A_{78}=0$,
and
\begin{eqnarray}
\partial_-(\alpha^m_1+\alpha^m_2+\alpha^m_3)=0,
\label{partial1'}
\end{eqnarray}
because, from (\ref{alpha}), $\alpha_i^{2n}=\alpha_i^{2n-1}$
while $\mu_n^{2n}=-\mu_n^{2n-1}$.
Consequently $A_m$ must be independent of $x^-$,
because the right hand
side of (\ref{A_m1}) is independent of $x^-$
from eq.~(\ref{partial1'}).
Noting that $\alpha_i^m$ is related to $a_i$ in (\ref{alpha}),
we find that eq.~(\ref{partial1'})
leads to four differential equations
for $a_i$,
\begin{eqnarray}
&&
\partial_-(a_1-a_2-a_3)=0,~~~
\partial_-(-a_1+a_2-a_3)=0,\nonumber\\&&
\partial_-(-a_1-a_2+a_3)=0,~~~
\partial_-(a_1+a_2+a_3)=0,
\end{eqnarray}
which imply
\begin{eqnarray}
\partial_-a_1=\partial_-a_2=\partial_-a_3=0.
\end{eqnarray}
This means that $\xi$ is independent of $x^-$.
The extra Killing spinors are determined as a non-trivial solution of
\begin{eqnarray}
&
\Bigl[&
\Big(
-4(\alpha^m_1+\alpha^m_2+\alpha^m_3
)^2
-A_m
\Big)~\1~
+
16\alpha^m_1(\alpha^m_2+\alpha^m_3)
~(\PP_3
+\PP_{14})
\nonumber\\&&
+16\alpha^m_3(\alpha^m_1+\alpha^m_2)
~(\PP_5
+\PP_{12})~
+
16\alpha^m_2(\alpha^m_1+\alpha^m_3)
\PP_8+\PP_9
)
~~
\Bigr]\varepsilon_-=0.
\label{reduced KSE1}
\end{eqnarray}
which reveals the four-fold degeneracy of the extra Killing
spinors. If (\ref{A_m1}) is satisfied,
$\PP_2\varepsilon_-$ and $\PP_{15}\varepsilon_-$
are a pair of complex extra Killing spinors and the background
admits 20 Killing spinors, 16 standard and 4 extra Killing spinors.
If, in addition, the coefficient of $(\PP_I+\PP_{17-I})$, $I=3,5,8$,
vanishes,
then 
$\PP_I\varepsilon_-$,
and $\PP_{17-I}\varepsilon_-$
give a pair of additional complex extra Killing spinors.

The backgrounds with extra supersymmetries
automatically satisfy the supergravity equation of motion (\ref{sugra})
because (\ref{A_m1}) and (\ref{xis1}) lead to
\begin{eqnarray}
&&\triangle H=\sum_{m=1}^82A_m=-64(a_1^2+a_2^2+a_3^2),\\
&&-\frac{32}{4!}\xi_{mnpq}\xi^{mnpq}=-64(a_1^2+a_2^2+a_3^2).
\end{eqnarray}

\subsection{the case~\p{xim2}}

In this case, all the matrices (\ref{matrices-2}) can be
constructed as products of four matrices, $\gamma^{2367}$,
$\gamma^{1256}$,
$\gamma^{1247}$ and $\gamma^{12...8}$. From these matrices, we make
four rank-8 projection operators
\begin{eqnarray}
&&P_1=\frac{1}{2}(\1+\gamma^{2367}),~~~
P_2=\frac{1}{2}(\1+\gamma^{1256}),
\nonumber\\&&
P_3=\frac{1}{2}(\1+\gamma^{1247}),~~~
P_4=\frac{1}{2}(\1+\gamma^{12...8}).
\label{P1'}
\end{eqnarray}
We rewrite (\ref{KSE}) in terms of these projection operators.
Because $\gamma^{12..8}\varepsilon_-=-\varepsilon_-$,
one finds that
\begin{eqnarray}
\xis_{(m)}\varepsilon_-&=&
\Big(~~
   2\beta_1^m \gamma^{4567}
 + 2\beta_2^m \gamma^{2367}
 + 2\beta_3^m \gamma^{2345}
 + 2\beta_4^m \gamma^{1357}
 \nonumber\\&&~~~
 + 2\beta_5^m \gamma^{1346}
 + 2\beta_6^m \gamma^{1256}
 + 2\beta_7^m \gamma^{1247}
~~~\Big)\varepsilon_-.
\label{xim on e}
\end{eqnarray}
Because
\begin{eqnarray}
&&
\gamma^{4567}= -(2P_2-\1)(2P_3-\1),~~~
\gamma^{2367}=(2P_1-\1),
\nonumber\\&&
\gamma^{2345}=(2P_1-\1)(2P_2-\1)(2P_3-\1),~~~
\gamma^{1357}=(2P_1-\1)(2P_2-\1),
\nonumber\\&&
\gamma^{1346}=-(2P_1-\1)(2P_3-\1),~~~
\gamma^{1256}=(2P_2-\1),~~~
\gamma^{1247}=(2P_3-\1)
\end{eqnarray}
$\xis_{(m)}\varepsilon_-$ in (\ref{xim on e})
can be rewritten as
\begin{eqnarray}
\xis_{(m)}\varepsilon_-&=&\Big[~
-2(\beta^m_1
+\beta^m_2
+\beta^m_3
-\beta^m_4
+\beta^m_5
+\beta^m_6
+\beta^m_7
)~\1
\nonumber\\&&~~
+4(\beta^m_2
 +\beta^m_3
 -\beta^m_4
 +\beta^m_5
)~P_1
+4(\beta^m_1
 +\beta^m_3
 -\beta^m_4
 +\beta^m_6
)~P_2
\nonumber\\&&~~
+4(\beta^m_1
 +\beta^m_3
 +\beta^m_5
 +\beta^m_7
)~P_3
-8(
 \beta^m_3
 -\beta^m_4
)~P_1P_2
-8(
 \beta^m_3
 +\beta^m_5
)~P_1P_3
\nonumber\\&&~~
-8(
 \beta^m_1
 +\beta^m_3
)~P_2P_3
+16\beta^m_3
~P_1P_2P_3
~~\Big]\varepsilon_-.
\end{eqnarray}
Substituting this into (\ref{KSE}) yields
\begin{eqnarray}
&\Bigl[&
\Big(
-2i\partial_-(
\beta^m_1+\beta^m_2+\beta^m_3-\beta^m_4
+\beta^m_5+\beta^m_6+\beta^m_7
)
\nonumber\\&&~~~~
-4(
\beta^m_1+\beta^m_2+\beta^m_3-\beta^m_4
+\beta^m_5+\beta^m_6+\beta^m_7
)^2
-A_m
\Big)~\1
\nonumber\\&&
+\Big(
4i\partial_-(
\beta^m_2+\beta^m_3-\beta^m_4+\beta^m_5
)
+16(\beta^m_2+\beta^m_3-\beta^m_4+\beta^m_5)
 (\beta^m_1+\beta^m_6+\beta^m_7)
\Big)~P_1
\nonumber\\&&
+\Big(
4i\partial_-
(
 \beta^m_1+\beta^m_3+\beta^m_6-\beta^m_4
)
+16
 (\beta^m_1+\beta^m_3+\beta^m_6-\beta^m_4 )
 (\beta^m_2+\beta^m_5+\beta^m_7)
\Big)~P_2
\nonumber\\&&
+\Big(
4i\partial_-(\beta^m_1+\beta^m_3+\beta^m_5+\beta^m_7)
 +16(\beta^m_1+\beta^m_3+\beta^m_5+\beta^m_7)
 (\beta^m_2-\beta^m_4+\beta^m_6)
\Big)~P_3
\nonumber\\&&
+\Big(
-8i\partial_-(
\beta^m_3-\beta^m_4
)
-32(\beta^m_3-\beta^m_4)\beta^m_7
-32(\beta^m_1+\beta^m_6)(\beta^m_2+\beta^m_5)
\Big)~P_1P_2
\nonumber\\&&
+\Big(
-8i\partial_-(
\beta^m_3+\beta^m_5
)
-32(
\beta^m_3+\beta^m_5
)\beta^m_6
-32(\beta^m_1+\beta^m_7)\beta^m_2-\beta^m_4)
\Big)~P_3P_1
\nonumber\\&&
+\Big(
-8i\partial_-(
\beta^m_1+\beta^m_3
)
-32(
\beta^m_1+\beta^m_3
)\beta^m_2
-32(-\beta^m_4+\beta^m_6)\beta^m_5-\beta^m_7)
\Big)~P_3P_2
\nonumber\\&&
+\Big(
16i\partial_-\beta^m_3
+64(\beta^m_1\beta^m_2
+\beta^m_5\beta^m_6
-\beta^m_4\beta^m_7)
\Big)~P_1P_2P_3
~~\Big]\varepsilon_-=0.
\label{KSE2}
\end{eqnarray}
Introducing rank-1 projection operators of a 16-component
spinor onto the $I$-th component,
\begin{eqnarray}
\PP_I=diag(0,\ldots,0,\stackrel{I}{1},0,\ldots,0),
\end{eqnarray}
the rank-8 projection operators $P_A$ can be expressed in terms of
these rank-1 projection operators as
\begin{eqnarray}
&&P_1=\sum_{I=1,2,..,8}\PP_I,~~~
P_2=\sum_{I=1,2,3,4,9,10,11,12}\PP_I,\nonumber\\&&
P_3=\sum_{I=1,2,5,6,9,10,13,14}\PP_I,~~~
P_4=\sum_{I=1,3,5,7,9,11,13,15}\PP_I.
\label{PP1'}
\end{eqnarray}
Noting that
non-trivial spinors are $\PP_I\varepsilon_-$
with $I=2,4,6,8,10,12,14$ and $16$,
because the chirality condition is
\begin{eqnarray}
\frac{1}{2}(\gamma^{12...8}+\1)\varepsilon_-=
P_4\varepsilon_-=0,
\end{eqnarray}
(\ref{KSE2}) reduces to
\begin{eqnarray}
&
\Bigl[&
\Big(
-2i\partial_-(
\beta^m_1+\beta^m_2+\beta^m_3-\beta^m_4+\beta^m_5+\beta^m_6+\beta^m_7
)
\nonumber\\&&~~~~
-4(
\beta^m_1+\beta^m_2+\beta^m_3-\beta^m_4+\beta^m_5+\beta^m_6+\beta^m_7
)^2
-A_m
\Big)~\1
\nonumber\\&&
+\Big(
4\partial_-(\beta^m_2+\beta^m_3+\beta^m_6+\beta^m_7)
+16(\beta^m_2+\beta^m_3+\beta^m_6+\beta^m_7)
 (\beta^m_1-\beta^m_4+\beta^m_5)
\Big)~
\PP_2
\nonumber\\&& 
+\Big(
4\partial_-(\beta^m_1+\beta^m_2+\beta^m_5+\beta^m_6)
+16(\beta^m_1+\beta^m_2+\beta^m_5+\beta^m_6)
(\beta^m_3-\beta^m_4+\beta^m_7)
\Big)~\PP_4
\nonumber\\&& 
+\Big(
4\partial_-(\beta^m_1+\beta^m_2-\beta^m_4+\beta^m_7)
+16(\beta^m_1+\beta^m_2-\beta^m_4+\beta^m_7)
 (\beta^m_3+\beta^m_5+\beta^m_6)
\Big)~\PP_6
\nonumber\\&& 
+\Big(
4\partial_-(\beta^m_2+\beta^m_3-\beta^m_4+\beta^m_5)
+16(\beta^m_2+\beta^m_3-\beta^m_4+\beta^m_5)
   (\beta^m_1+\beta^m_6+\beta^m_7)
\Big)~\PP_8
\nonumber\\&& 
+\Big(
4\partial_-(-\beta^m_4+\beta^m_5+\beta^m_6+\beta^m_7)
+16(-\beta^m_4+\beta^m_5+\beta^m_6+\beta^m_7)
   (\beta^m_1+\beta^m_2+\beta^m_3)
\Big)~\PP_{10}
\nonumber\\&& 
+\Big(
4\partial_-(\beta^m_1+\beta^m_3-\beta^m_4+\beta^m_6)
+16(\beta^m_1+\beta^m_3-\beta^m_4+\beta^m_6)
   (\beta^m_2+\beta^m_5+\beta^m_7)
\Big)~\PP_{12}
\nonumber\\&& 
+\Big(
4\partial_-(\beta^m_1+\beta^m_3+\beta^m_5+\beta^m_7)
+16(\beta^m_1+\beta^m_3+\beta^m_5+\beta^m_7)
   (\beta^m_2-\beta^m_4+\beta^m_6)
\Big)~\PP_{14}
\nonumber\\&&\hspace{100mm}
\Bigr]\varepsilon_-=0.
\end{eqnarray}
Again the coefficient of $\1$ in this equation
must vanish in order to give a complex extra Killing spinor.
Because $\beta_i^m$ and $A_m$ are real, this implies that
\begin{eqnarray}
&&\partial_-(\beta^m_1+\beta^m_2+\beta^m_3-\beta^m_4
 +\beta^m_5+\beta^m_6+\beta^m_7)=0,
\label{partial}\\
&&A_m=-4(\beta^m_1+\beta^m_2+\beta^m_3-\beta^m_4
 +\beta^m_5+\beta^m_6+\beta^m_7)^2.
\label{A_m}
\end{eqnarray}
Consequently $A_m$ must be independent of $x^-$, because the right hand
side of (\ref{A_m}) is independent of $x^-$ from eq.~(\ref{partial}).
Noting that $\beta_i^m$ is related to $b_i$ in (\ref{beta}),
we find that eq.~(\ref{partial}) leads to four differential equations
for $b_i$
which imply
\begin{eqnarray}
\partial_-b_1=\partial_-b_2=...=\partial_-b_7=0,
\end{eqnarray}
which implies that $\xi$ is independent of $x^-$.
The extra Killing spinors are determined as a non-trivial solution of
\begin{eqnarray}
&
\Bigl[&
\Big(
-4(
\beta^m_1+\beta^m_2+\beta^m_3-\beta^m_4+\beta^m_5+\beta^m_6+\beta^m_7
)^2
-A_m
\Big)~\1
\nonumber\\&&
+
16(\beta^m_2+\beta^m_3+\beta^m_6+\beta^m_7)
 (\beta^m_1-\beta^m_4+\beta^m_5)
~\PP_2
\nonumber\\&& 
+16(\beta^m_1+\beta^m_2+\beta^m_5+\beta^m_6)
(\beta^m_3-\beta^m_4+\beta^m_7)
~\PP_4
\nonumber\\&& 
+16(\beta^m_1+\beta^m_2-\beta^m_4+\beta^m_7)
 (\beta^m_3+\beta^m_5+\beta^m_6)
~\PP_6
\nonumber\\&& 
+16(\beta^m_2+\beta^m_3-\beta^m_4+\beta^m_5)
   (\beta^m_1+\beta^m_6+\beta^m_7)
~\PP_8
\nonumber\\&& 
+16(-\beta^m_4+\beta^m_5+\beta^m_6+\beta^m_7)
   (\beta^m_1+\beta^m_2+\beta^m_3)
~\PP_{10}
\nonumber\\&& 
+16(\beta^m_1+\beta^m_3-\beta^m_4+\beta^m_6)
   (\beta^m_2+\beta^m_5+\beta^m_7)
~\PP_{12}
\nonumber\\&& 
+16(\beta^m_1+\beta^m_3+\beta^m_5+\beta^m_7)
   (\beta^m_2-\beta^m_4+\beta^m_6)
~\PP_{14}
~~~
\Bigr]\varepsilon_-=0.
\label{reduced KSE2}
\end{eqnarray}
which again shows the two-fold degeneracy of the extra Killing
spinors. If (\ref{A_m}) is satisfied,
$\PP_{16}\varepsilon_-$
is a complex extra Killing spinor and the background
admits 18 Killing spinors, 16 standard and 2 extra Killing spinors.
If, in addition, the coefficient of $\PP_I$, $I=2,4,6,8,10,12,14$,
is zero,
then 
$\PP_{I}\varepsilon_-$
becomes an additional complex extra Killing spinor.

Again, the backgrounds with extra supersymmetries
automatically satisfy the supergravity equation of motion
(\ref{sugra})
because (\ref{A_m}) and (\ref{xis2}) lead to
\begin{eqnarray}
&&\triangle H=\sum_{m=1}^82A_m
 =-64(b_1^2+b_2^2+b_3^2+b_4^2+b_5^2+b_6^2+b_7^2),\\
&&-\frac{32}{4!}\xi_{mnpq}\xi^{mnpq}
 =-64(b_1^2+b_2^2+b_3^2+b_4^2+b_5^2+b_6^2+b_7^2).
\end{eqnarray}

\bigskip

In summary, we have shown that
IIB pp-wave backgrounds
with a self-dual five-form R-R flux
can be reduced to the form modulo coordinate transformations
\begin{eqnarray}
ds^2=
 2dx^+dx^- +A_{m}x^mx^m(dx^-)^2+(dx^m)^2,~~~
F=dx^-\wedge \xi,
\label{pp}
\end{eqnarray}
where $A_m$ and $\xi$ are constants,
if the backgrounds admit extra Killing spinors
characterized by (\ref{matrices-1}) and (\ref{matrices-2}).
It is interesting to examine the cases in which
the extra Killing spinors are characterized
by more general Cartan matrices.

\sect{IIB pp-wave backgrounds with extra supersymmetries}
In the previous section, we have shown that
Killing spinor equations reduce to
(\ref{reduced KSE1}) and (\ref{reduced KSE2}).
The former case admits NS-NS and R-R three-forms
in addition to a self-dual R-R five-form
and has been examined in the literature,
whereas the latter case does not admit them
and has not been examined well.
We provide pp-wave backgrounds
obtained by solving (\ref{reduced KSE1}) and
(\ref{reduced KSE2})
in this section.

For the former case, (\ref{reduced KSE1}),
one finds three classes of solutions.
One is the maximally supersymmetric pp-wave background found
in \cite{BFHP:A new maximally}
\begin{eqnarray}
A_m=-4\mu^2,~~m=1,...,8,~~~~
\xis=\mu(\gamma^{1234}+\gamma^{5678}),
\label{1-1}
\end{eqnarray}
which means
\begin{eqnarray}
\xi=\mu(dx^1\wedge dx^2\wedge dx^3\wedge dx^4
+dx^5\wedge dx^6\wedge dx^7\wedge dx^8)
\end{eqnarray}
because $\xis=\frac{1}{4!}\xi_{lmnp}\gamma^{lmnp}$.
We will use $\xis$ below to indicate $\xi$
for compact expressions.
Another is the background with 24 supersymmetries
\begin{eqnarray}
A_m=\left\{
  \begin{array}{ll}
  -4(\mu_1-\mu_2)^2,     &m=1,2,5,6,    \\
  -4(\mu_1+\mu_2)^2,     &m=3,4,7,8,    \\
  \end{array}
\right.
~~~
\xis=\mu_1(\gamma^{1234}+\gamma^{5678})
+\mu_2(\gamma^{3456}+\gamma^{1278}),
\label{1-2}
\end{eqnarray}
which is a subclass of solutions found in \cite{BR:Supergravity}.
The third is the 20 supersymmetric background
\begin{eqnarray}
A_m&=&
\left\{
  \begin{array}{ll}
-4(-\mu_1+\mu_2-\mu_3)^2,  &m=1,2,    \\
-4(-\mu_1-\mu_2+\mu_3)^2,  &m=3,4,    \\
-4(\mu_1-\mu_2-\mu_3)^2,   &m=5,6,    \\
-4(\mu_1+\mu_2+\mu_3)^2,   &m=7,8,    \\
  \end{array}
\right.\nonumber\\
\xis&=& \mu_1(\gamma^{1234}+\gamma^{5678})
 +\mu_2(\gamma^{3456}+\gamma^{1278})
 +\mu_3(\gamma^{1256}+\gamma^{3478}),
\end{eqnarray}
which contains the above two backgrounds,
(\ref{1-1}) and (\ref{1-2}),
 as special cases.

\medskip

For the latter case, (\ref{reduced KSE2}),
we find IIB pp-wave solutions, which have not been given
in the literature
as long as we know.
Eqn. (\ref{reduced KSE2}) leads to
two maximally supersymmetric pp-wave backgrounds.
One is the same as (\ref{1-1}),
and the other is
\begin{eqnarray}
A_m&=&-4\mu^2,~~m=1,...,8,
\label{2-2}\\
\xis&=&
\frac{\mu}{2}\Big(
 -(\gamma^{4567}+\gamma^{1238})
 +(\gamma^{2367}+\gamma^{1458})
 +(\gamma^{1346}+\gamma^{2578})
 +(\gamma^{1256}+\gamma^{3478})
\Big).
\nonumber
\end{eqnarray}
The self-dual four-form can be regarded as
the real or imaginary part
of the holomorphic $(4,0)$-form
of a Calabi-Yau four-fold with $SU(4)$ holonomy.
It can be shown that there are twelve Lorentz generators
in this background, which is the same as
in the $SO(4)\times SO(4)$ case (\ref{1-1}).
In addition,
$\xis$ in (\ref{2-2}) satisfies the Pl\"ucker-type relation
and thus, as proven in \cite{Plucker},
$\xis$ must decompose into orthogonal pieces like $\xis$
in (\ref{1-1}).
These suggest that
(\ref{2-2}) can be related to (\ref{1-1})
by a coordinate transformation.
We show that this is the case\footnote{
We thank Jos\'e Figueroa-O'Farrill
for correspondence.
}.
Because $\xis$ in (\ref{2-2}) can be rewritten as
\begin{eqnarray}
\xis=\frac{\mu}{4}\Big(
(\gamma^1-\gamma^7)(\gamma^2-\gamma^4)(\gamma^5+\gamma^3)(\gamma^6-\gamma^8)
+
(\gamma^1+\gamma^7)(\gamma^2+\gamma^4)(\gamma^5-\gamma^3)(\gamma^6+\gamma^8)
\Big),
\end{eqnarray}
$\xis$ in (\ref{2-2}) is transformed to $\xis$ in (\ref{1-1})
by the coordinate transformation
\begin{eqnarray}
y^{1,5}=\frac{1}{\sqrt{2}}(x^1\mp x^7),~
y^{2,6}=\frac{1}{\sqrt{2}}(x^2\mp x^4),~
y^{3,7}=\frac{1}{\sqrt{2}}(x^5\pm x^3),~
y^{4,8}=\frac{1}{\sqrt{2}}(x^6\mp x^8).
\end{eqnarray}
This transformation does not change the metric,
and thus we have rewritten (\ref{2-2}) as (\ref{1-1}).

As non-maximally supersymmetric backgrounds,
we find pp-wave solutions with 24, 20 and 18 supersymmetries.
The background
\begin{eqnarray}
A_m&=&
\left\{
  \begin{array}{ll}
-64\mu^2,      &m=1,3,5,7,    \\
-16\mu^2,       &m=2,4,6,8,    \\
  \end{array}
\right.
\nonumber\\
\xis&=&
  \mu\Big(
 -(\gamma^{4567}+\gamma^{1238})
 +(\gamma^{2367}+\gamma^{1458})
 -(\gamma^{2345}+\gamma^{1678})
 -2(\gamma^{1357}+\gamma^{2468})
 \nonumber\\&&~~~~~~
 +(\gamma^{1346}+\gamma^{2578})
 -(\gamma^{1256}+\gamma^{3478})
 +(\gamma^{1247}+\gamma^{3568})
\Big),
\label{2-3}
\end{eqnarray}
admits 24 supersymmetries.
We find three classes of 20 supersymmetric backgrounds,
a four-parameter family
\begin{eqnarray}
A_m&=&
\left\{
  \begin{array}{ll}
-4(\mu_1+\mu_2-\mu_3-\mu_4)^2, &m=1,3,    \\
-4(\mu_1+\mu_2+\mu_3+\mu_4)^2 ,&m=2,8,    \\
-4(\mu_1-\mu_2+\mu_3-\mu_4)^2, &m=4,6,    \\
-4(\mu_1-\mu_2-\mu_3+\mu_4)^2 ,&m=5,7,    \\
  \end{array}
\right.
\label{2-4}\\
\xis&=&
 \mu_1(\gamma^{2367}+\gamma^{1458})
 +\mu_2(\gamma^{2345}+\gamma^{1678})
 +\mu_3(\gamma^{1256}+\gamma^{3478})
 +\mu_4(\gamma^{1247}+\gamma^{3568})
 \nonumber
\end{eqnarray}
and two three parameter families,
\begin{eqnarray}
A_m&=&
\left\{
  \begin{array}{ll}
-4(2\mu_1+2\mu_2)^2,    &m=1,3,    \\
-16\mu_3^2 ,      &m=2,5,7,8,    \\
-4(2\mu_1-2\mu_2)^2 ,      & m=4,6,   \\
  \end{array}
\right.
\nonumber\\
\xis&=&
 -\mu_3(\gamma^{4567}+\gamma^{1238})
 +\mu_1(\gamma^{2367}+\gamma^{1458})
 +\mu_2(\gamma^{2345}+\gamma^{1678})
 \nonumber\\&&
 +\mu_3(\gamma^{1357}+\gamma^{2468})
 -\mu_1(\gamma^{1247}+\gamma^{3568})
 -\mu_2(\gamma^{1256}+\gamma^{3478})
\label{2-5}
\end{eqnarray}
and
\begin{eqnarray}
A_m&=&
\left\{
  \begin{array}{ll}
-64\mu_1^2       ,&m=1,3,    \\
-16(-\mu_2+\mu_3)^2      , &m=2,8,    \\
 -16\mu_3^2    ,   &m=4,6,    \\
-16\mu_2^2     ,  &m=5,7,    \\
  \end{array}
\right.
\nonumber\\
\xis&=&
 -\mu_2(\gamma^{4567}+\gamma^{1238})
 -\mu_1(\gamma^{2367}+\gamma^{1458})
 -\mu_1(\gamma^{2345}+\gamma^{1678})
 +\mu_2(\gamma^{1357}+\gamma^{2468}) \nonumber\\&&
 +\mu_3(\gamma^{1346}+\gamma^{2578})
 +\mu_1(\gamma^{1256}+\gamma^{3478})
 +\mu_1(\gamma^{1247}+\gamma^{3568}).
\label{2-6}
\end{eqnarray}
These backgrounds can be obtained as special cases
of the backgrounds with 18 Killing spinors 
\begin{eqnarray}
A_m&=&\Big(
-4(\mu_1+\mu_2+\mu_3+\mu_4-\mu_5-\mu_6-\mu_7)^2, 
-4(\mu_1-\mu_2-\mu_3-\mu_4+\mu_5-\mu_6-\mu_7)^2, 
\nonumber\\&&
-4(\mu_1-\mu_2-\mu_3+\mu_4-\mu_5+\mu_6+\mu_7)^2, 
-4(-\mu_1+\mu_2-\mu_3-\mu_4-\mu_5+\mu_6-\mu_7)^2, 
\nonumber\\&&
-4(-\mu_1+\mu_2-\mu_3+\mu_4+\mu_5-\mu_6+\mu_7)^2, 
-4(-\mu_1-\mu_2+\mu_3-\mu_4-\mu_5-\mu_6+\mu_7)^2, 
\nonumber\\&&
-4(-\mu_1-\mu_2+\mu_3+\mu_4+\mu_5+\mu_6-\mu_7)^2, 
-4(\mu_1+\mu_2+\mu_3-\mu_4+\mu_5+\mu_6+\mu_7)^2
\Big),
\nonumber\\
\xis&=&  \mu_1(\gamma^{4567}+\gamma^{1238})
 +\mu_2(\gamma^{2367}+\gamma^{1458})
 +\mu_3(\gamma^{2345}+\gamma^{1678})
 +\mu_4(\gamma^{1357}+\gamma^{2468})
 \nonumber\\&&
 +\mu_5(\gamma^{1346}+\gamma^{2578})
 +\mu_6(\gamma^{1256}+\gamma^{3478})
 +\mu_7(\gamma^{1247}+\gamma^{3568}).
\end{eqnarray}

\sect{Summary and Discussions}
We have established a uniqueness theorem which states that
any IIB pp-wave background of the form (\ref{background})
can be reduced to the form~(\ref{pp1}) modulo
coordinate transformations,
if there exist at least one non-harmonic extra Killing spinor.
We examined further the cases in which extra Killing spinors
are characterized by (\ref{matrices-1}) and (\ref{matrices-2}).
Examining Killing spinor equations,
we found IIB pp-wave backgrounds
which admit 18, 20, 24 and 32 Killing spinors.

It is interesting to examine the similar uniqueness theorem
for pp-wave backgrounds 
of IIA supergravity and supergravities in lower dimensions.
We expect that the similar uniqueness theorems
can be established.

We have seen that two sets of mutually commuting matrices
(\ref{matrices-1}) and (\ref{matrices-2})
can be related to
the K\"ahler form $J$ of a Calabi-Yau four-fold with $SU(4)$ holonomy 
and the self-dual Cayley four-form $\Psi$
of $d=8$ Riemannian manifold with $Spin(7)$ holonomy, respectively.
For eleven-dimensions \cite{OS},
two sets of mutually commuting matrices were shown to be
related to
the K\"ahler form $J$ of a Calabi-Yau four-fold with $SU(4)$ holonomy 
and the associative three-form
of $d=7$ Riemannian manifold with $G_2$ holonomy.
These may suggest that
Killing spinor equations for $d$-dimensional pp-wave backgrounds
with flux
can be reduced to
those for $(d-2)$-dimensional backgrounds without flux.
If this is the case,
a classification of pp-wave backgrounds with flux
gets simplified.
Further, it is interesting to try to construct $(d-2)$-dimensional
backgrounds with special holonomy
from $d$-dimensional pp-wave backgrounds with flux.

We have seen that
the pp-wave backgrounds which admit at least one extra Killing spinor
automatically satisfy the supergravity equation of motion,
which suggests that
Killing spinor equations for backgrounds with extra supersymmetries
have rich algebraic structures,
just as those for maximally supersymmetric backgrounds.
It is interesting to classify all non-maximally supersymmetric backgrounds
which admit more than 16 supersymmetries,
as was achieved for the maximally supersymmetric case in
\cite{FP;Maximally}.

\section*{Acknowledgements}
We thank Nobuyoshi Ohta and Kentaroh Yoshida for helpful discussions.



\begin{thebibliography}{1}
\bibitem{BFHP:A new maximally}
M.~Blau, J.~Figueroa-O'Farrill, C.~Hull and G.~Papadopoulos,
``A new maximally supersymmetric background
of IIB superstring theory,''
JHEP {\bf 0201} (2002) 047
[arXiv:hep-th/0110242].
\bibitem{Metsaev:Type IIB}
R.~R.~Metsaev,
``Type IIB Green-Schwarz superstring in plane wave
Ramond-Ramond  background,''
Nucl.\ Phys.\ B {\bf 625} (2002) 70
[arXiv:hep-th/0112044].
\bibitem{Penrose}
R.~Penrose,
``Any space-time has a plane wave as a limit'',
in Differential Geometry and Relativity, Cahen and Flato eds. (1976)
Reidel Publishing, Dordrecht-Holland.\\
R.~Gueven,
``Plane wave limits and T-duality,''
Phys.\ Lett.\ B {\bf 482} (2000) 255
[arXiv:hep-th/0005061].
\bibitem{BMN}
D.~Berenstein, J.~M.~Maldacena and H.~Nastase,
``Strings in flat space and pp waves from N = 4 super Yang Mills,''
JHEP {\bf 0204} (2002) 013
[arXiv:hep-th/0202021],
and references thereof.
\bibitem{BFCP:Penrose limits}
M.~Blau, J.~Figueroa-O'Farrill, C.~Hull and G.~Papadopoulos,
`Penrose limits and maximal supersymmetry,''
Class.\ Quant.\ Grav.\  {\bf 19} (2002) L87
[arXiv:hep-th/0201081].
\bibitem{HKS:IIB}
M.~Hatsuda, K.~Kamimura and M.~Sakaguchi,
``From super-AdS$_5\times S^5$ algebra
to super-pp-wave algebra,''
Nucl.\ Phys.\ B {\bf 632} (2002) 114
[arXiv:hep-th/0202190].
\bibitem{CLP:Penrose}
M.~Cvetic, H.~Lu and C.~N.~Pope,
``Penrose limits, pp-waves and deformed M2-branes,''
arXiv:hep-th/0203082.
\bibitem{CLP:M-theory pp-waves}
M.~Cvetic, H.~Lu and C.~N.~Pope,
``M-theory pp-waves, Penrose limits
and supernumerary supersymmetries,''
Nucl.\ Phys.\ B {\bf 644}, 65 (2002)
[arXiv:hep-th/0203229].
\bibitem{CHKW:Penrose limit of RG fixed points}
R.~Corrado, N.~Halmagyi, K.~D.~Kennaway and N.~P.~Warner,
``Penrose limits of RG fixed points and pp-waves with background fluxes,''
Adv.\ Theor.\ Math.\ Phys.\  {\bf 6} (2003) 597
[arXiv:hep-th/0205314].
\bibitem{BJLM:penrose limits deformed pp-waves}
D.~Brecher, C.~V.~Johnson, K.~J.~Lovis and R.~C.~Myers,
``Penrose limits, deformed pp-waves and the string duals of $N = 1$
large $N$  gauge theory,''
JHEP {\bf 0210} (2002) 008
[arXiv:hep-th/0206045].
\bibitem{BR:Supergravity}
I.~Bena and R.~Roiban,
``Supergravity pp-wave solutions with 28 and 24 supercharges,''
Phys.\ Rev.\ D {\bf 67} (2003) 125014
[arXiv:hep-th/0206195].
\bibitem{GPS:penrose limit and RG flow}
E.~G.~Gimon, L.~A.~Pando Zayas and J.~Sonnenschein,
``Penrose limits and RG flows,''
arXiv:hep-th/0206033.
\bibitem{Kowalski-Glikman:vacuum}
J.~Kowalski-Glikman,
``Vacuum states in supersymmetric Kaluza-Klein theory,''
Phys.\ Lett.\ B {\bf 134}, 194 (1984).
\bibitem{Chrusciel:The isometry}
P.~T.~Chrusciel and J.~Kowalski-Glikman,
``The isometry group and Killing spinors for the pp wave space-time in
$D = 11$ supergravity,''
Phys.\ Lett.\ B {\bf 149}, 107 (1984).
\bibitem{Figueroa-O'Farrill:Homogeneous}
J.~Figueroa-O'Farrill and G.~Papadopoulos,
``Homogeneous fluxes, branes and a maximally supersymmetric solution of
M-theory,''
JHEP {\bf 0108}, 036 (2001)
[arXiv:hep-th/0105308].
\bibitem{HKS:Super-pp-wave}
M.~Hatsuda, K.~Kamimura and M.~Sakaguchi,
``Super-pp-wave algebra from super-AdS $\times$ S algebras in
eleven-dimensions,''
Nucl.\ Phys.\ B {\bf 637} (2002) 168
[arXiv:hep-th/0204002].
\bibitem{Michelson:Twisted}
J.~Michelson,
``(Twisted) toroidal compactification of pp-waves,''
Phys.\ Rev.\ D {\bf 66} (2002) 066002
[arXiv:hep-th/0203140].
\bibitem{GH:pp-waves in 11-dimensions}
J.~P.~Gauntlett and C.~M.~Hull,
``pp-waves in 11-dimensions with extra supersymmetry,''
JHEP {\bf 0206} (2002) 013
[arXiv:hep-th/0203255].
\bibitem{Michelson:A pp-wave}
J.~Michelson,
``A pp-wave with 26 supercharges'',
Class.\ Quant.\ Grav.\ {\bf 19} (2002) 5935-5949,
hep-th/0206204.
\bibitem{FP;Maximally}
J.~Figueroa-O'Farrill and G.~Papadopoulos,
``Maximally supersymmetric solutions of ten-dimensional
and eleven-dimensional supergravities,''
JHEP {\bf 0303} (2003) 048
[arXiv:hep-th/0211089].
\bibitem{SY:IIA}
K.~Sugiyama and K.~Yoshida,
``Type IIA string and matrix string on pp-wave,''
Nucl.\ Phys.\ B {\bf 644} (2002) 128
[arXiv:hep-th/0208029],\\
%
S.~j.~Hyun and H.~j.~Shin,
``N = (4,4) type IIA string theory on pp-wave background,''
JHEP {\bf 0210} (2002) 070
[arXiv:hep-th/0208074].
\bibitem{SS:String}
D.~Sadri and M.~M.~Sheikh-Jabbari,
``String theory on parallelizable pp-waves,''
JHEP {\bf 0306} (2003) 005
[arXiv:hep-th/0304169].
\bibitem{Meessen:A small note}
P.~Meessen,
``A small note on pp-wave vacua in 6 and 5 dimensions,''
Phys.\ Rev.\ D {\bf 65} (2002) 087501
[arXiv:hep-th/0111031].
\bibitem{Kowalski-Glikman:Positive}
J.~Kowalski-Glikman,
``Positive energy theorem and vacuum states
for the Einstein-Maxwell system,''
Phys.\ Lett.\ B {\bf 150} (1985) 125.
\bibitem{OS}
N.~Ohta and M.~Sakaguchi,
``Uniqueness of M-theory PP-Wave Background
with Extra Supersymmetries,''
arXiv:hep-th/0305215.
\bibitem{MM:Strings}
J.~Maldacena and L.~Maoz,
``Strings on pp-waves and massive two dimensional field theories,''
JHEP {\bf 0212} (2002) 046
[arXiv:hep-th/0207284].
\bibitem{BMO}
M.~Blau, P.~Meessen and M.~O'Loughlin,
``Goedel, Penrose, anti-Mach:
extra supersymmetries of time-dependent plane waves'',
arXiv:hep-th/0306161.
\bibitem{Wang}
M.~Y.~Wang,
``Parallel spinors and parallel forms,''
Ann. Global Anal. Geom. \textbf{7} (1989), 59-68.
\bibitem{Plucker}
J.~Figueroa-O'Farrill and G.~Papadopoulos,
``Pluecker-type relations for orthogonal planes,''
arXiv:math.ag/0211170.
\end{thebibliography}
\end{document}